\documentclass[pre,a4paper, twocolumn]{revtex4}
\usepackage{graphicx}
\usepackage{amsmath}
\usepackage{epsfig}
\usepackage{amssymb}
\usepackage{array}
\begin{document}

\title{Local modularity measure for network clusterizations}
\author{Stefanie Muff, Francesco Rao and Amedeo Caflisch}
\affiliation{Department of Biochemistry, University of Zurich,
             Winterthurerstrasse 190, CH-8057 Zurich, Switzerland\\
             tel: +41 1 635 55 21, fax: +41 1 635 68 62,
             e-mail: caflisch@bioc.unizh.ch}

\begin{abstract}
Many complex networks have an underlying modular structure, i.e., structural subunits (communities or clusters) characterized by highly interconnected nodes. The modularity $Q$ has been introduced as a measure to assess the quality of clusterizations. $Q$ has a global view, while in many real-world networks  clusters are linked mainly \emph{locally} among each other (\emph{local cluster-connectivity}). Here, we introduce a new measure, localized modularity $LQ$, which reflects local cluster structure. Optimization of $Q$ and $LQ$ on the clusterization of two biological networks shows that the localized modularity identifies more cohesive clusters, yielding a complementary view of higher granularity. 
 
\end{abstract} 

\date{\today}

\maketitle

Complex networks are a powerful tool for the analysis of a diverse range of systems including
technological \cite{Faloutsos, Albert:WWW}, social \cite{Scott:social,Newman:social}, and biological
networks \cite{Almaas:fluxes,Rao:protein}. Especially in biology, thanks to high-throughput experiments, there is a tremendous growth of available data that can be efficiently analyzed and summarized in terms of complex networks \cite{Xia,NetworkBiology}. In many cases, networks have an inherent modular structure which
can represent functional units, called communities or clusters, e.g., web pages of a certain subject \cite{Eckmann:WWW}, social groups \cite{Scott:social,Wasserman:social} or biological modules \cite{Amaral:cartography,Eisen:hierarchical}.
However, there is neither an obvious and commonly accepted definition of communities, nor a straightforward way to find the underlying modules of a network. 
Recently, many clustering algorithms have been proposed \cite{Girvan/Newman,Newman:fast,Reichardt:fuzzy,Radicchi,Enright,Palla:uncovering}. For a clusterization with $K$ communities, the \emph{modularity} $Q = \sum_{i=1}^K \left({e_{ii} -  (a_i)_{in} (a_i)_{out}} \right)$ has been introduced as a measure to assess the quality of a clusterization \cite{Newman:finding}, where  $e_{ii}=\frac{L_{i}}{L_{tot}}$, the effective fraction of links inside community $i$, is compared to $(a_i)_{in} (a_i)_{out}=\frac{(L_i)_{in} \cdot (L_i)_{out}}{L_{tot}^2}$ which is the predicted fraction of edges that fall into community $i$ if the links in a directed network are set between nodes without regard to the community structure. $Q$ is high when the clusterization is good and it can reach a maximum value of 1. Modularity is used to compare the quality of different clusterizations, e.g., to find the best split of a dendogram \cite{Newman:large} or to validate different clusterization methods and furthermore as fitness function in optimization procedures, where $Q_{max}$ should correspond to the objectively best clusterization of a network \cite{Newman:fast, Amaral:cartography}. The modularity is a global measure because the comparison of $\frac{L_{i}}{L_{tot}}$ with $\frac{(L_i)_{in} \cdot (L_i)_{out}}{L_{tot}^2}$ assumes that connections between all pairs of nodes are equally probable which reflects connectivity among all clusters.

On the other hand, in many complex networks most clusters are connected to only a small fraction of the remaining clusters. In metabolic networks, for instance, major pathways occur as clusters that are sparsely linked among each other \cite{Amaral:cartography}. Furthermore, in the protein folding network \cite{Rao:protein} communities are energy basins and transitions, i.e., connections, are allowed only between adjacent basins \cite{Reichardt:fuzzy}. We call this property \emph{local cluster-connectivity}. 
In this letter, we introduce a new measure for the quality of network clusterizations. 
%This yields a \emph{local} view of communities. 
To take into account local cluster-connectivity and overcome global network dependency, the approach of modularity is modified into a \emph{local} version. The contribution to modularity for each cluster $i$ is calculated for the subnetwork consisting of cluster $i$ and its neighbor clusters. This requires the determination of $i$'s neighborhood or, more precisely, all the links $L_{i_N}$ that are contained in this neighborhood. The sum of the contributions of all $K$ clusters yields
\begin{eqnarray*}
LQ = \sum_{i=1}^K \left({\frac{L_{i}}{L_{i_N}} -  \frac{(L_i)_{in} \cdot (L_i)_{out}}{(L_{i_N})^2}} \right) \ .
\end{eqnarray*}

\noindent We call $LQ$ \emph{localized modularity}. It is -- in contrast to $Q$ -- not bounded by 1, but can take any value. The more locally connected clusters a network has, the higher is $LQ$. On the other hand, in a network where all communities are linked among each other, $Q$ and $LQ$ coincide. 
\\

\begin{figure*}[th!]
\includegraphics[angle=0,width=16cm]{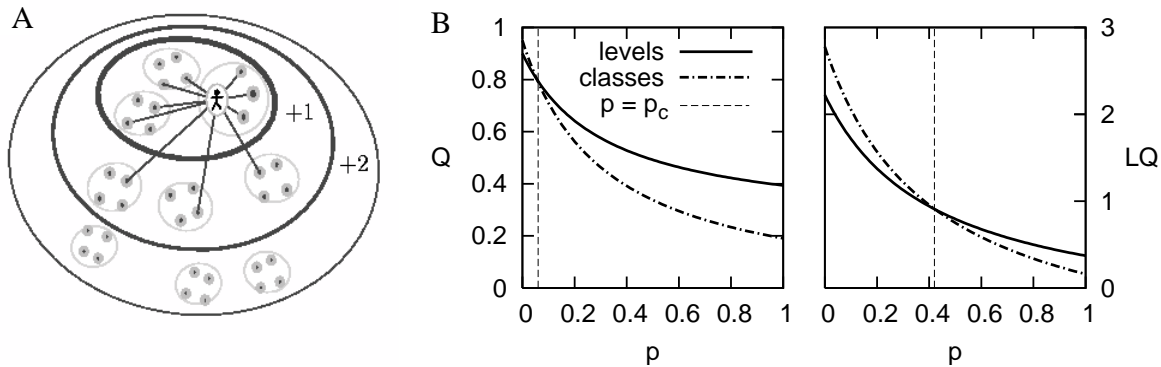}\\
\caption{\label{school} ({\bf A}) A student's view in the simplified schematic school network model with only 3 levels, 3 classes/level and 4 students/class: The student interacts with all his classmates, with other students on the same level with probability $p=0.5$ and with pupils one level above/below (+1) with probability $r=0.25$. No connections are assumed between students that are more than one level apart (+2 or more). ({\bf B}) The $p$-dependent behavior of the modularity and the localized modularity in 
the school network with 10 levels, 2 classes per level, 20 pupils per class and $r=\frac{p}{2}$. The modularity favors the
grouping of classes (solid line) in the same
level for almost all $p$, whereas localized modularity favors communities 
consisting of single classes (dot-dashed line) for $p<$0.42.}
\end{figure*}

It is interesting to compare the behavior of $Q$ and $LQ$ on different network topologies and use them as fitness functions for the optimization of network clusterizations \cite{Amaral:cartography,Newman:fast}. 
%Also the comparison of the $Q$-values of a network to the so-called \emph{null-case} of the corresponding random graph, as it was proposed by \cite{Guimera:random,Massen:communities}, can help to handle this. But if the modularity is used in an optimization procedure like in \cite{Newman:finding,Newman:fast,Massen:communities,Amaral:cartography}, the comparison to the null-case is not particularly relevant since one is only interested in the highest modularity $Q_{max}$. 
We start with an illustration of the differences between $Q$ and $LQ$ by discussing a simple example of a scalable local cluster-connectivity network, which we call the \emph{school network} (Fig.~\ref{school}A). It is a toy model of social interactions between pupils in a school with $l$ levels and $c$ classes per level. Levels have periodic boundary conditions to avoid spurious boundary effects (in the first and last level). In a real school, all the students of a class know each other and, as a first approximation, a student would interact most with people of his/her age. In the school network model, students are the nodes of the network and a link between two pupils is made if they know each other. Each class contains $s$ fully connected students. A link between two students of the same level but different classes is placed with a (high) probability $p\leq 1$ and connections between students that are one level
above/below (+1, Fig.~\ref{school}A) are made with smaller probability $r<p$. No social interaction is assumed between
persons that are more than one level apart from each other, i.e, if one of the students is more than one year older than the other (+2 or more, Fig.~\ref{school}A). Interestingly, when only two levels and two classes per level are considered, the school network model is essentially the same as the well-known (globally connected) 4 communities test network used in \cite{Newman:fast,Amaral:cartography}. Hence, the school network is a simple generalization to locally connected networks. It is unweighted and undirected but an extension to directed and weighted networks, e.g., asymmetrical friendship, is straightforward.

A grouping of all the pupils on one level into the same
cluster is reasonable for high $p$, i.e., when students of the same age interact among each other with high probability. But, as $p$ decreases, classes become more and more separated from each other until they fully break apart for $p=0$, where a fitness measure is expected to favor clusterizations that identify classes. Therefore, we calculated modularity and localized modularity
for the clusterization of nodes 
according to classes and according to levels for $p \in [0,1]$, $r=\frac{p}{2}$ and $s=20$ students per class. Figure~\ref{school}B shows the $Q$- and $LQ$-values for 10 levels and 2 classes per level. They were obtained analytically, using the expected numbers of links for each $p$. Both $Q$ and $LQ$ favor the clusterization into levels for $p$ close to 1. $LQ$ yields the same value for both clusterizations (\emph{crossing point}) at $p_c^{LQ}=0.42$ and prefers the
clusterization into classes for $p<0.42$. The modularity, on the other hand, has its crossing point at $p_c^Q=0.09$, i.e., it favors the classes only for $p<0.09$. In other words, $Q$ considers the classes and not the levels as the best cluster partition only if the probability of interaction between two students of the same age but different classes is smaller than 10$\%$.
%Note that a measure for the validation of clusterizations is expected to capture better the lowest-level, local community structure in the school network if the corresponding $p_c$ is high. This does not mean that one or another clusterization is right or wrong, it just shows that $Q$ and $LQ$ detect clusters at a different resolution.  
 
\begin{figure}[h!]
\begin{tabular}{c}
\includegraphics[angle=-90,width=7.5cm]{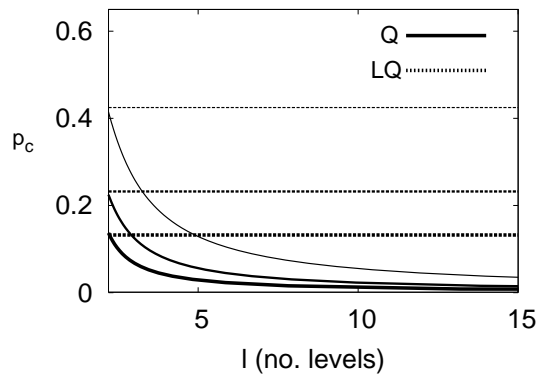}\\
\end{tabular}
\caption{\label{pc} Dependence of $p_c$ on network size: For 2, 5 and 10 classes/level (from top to bottom), $p_c^{LQ}$ (dotted lines) is always higher than $p_c^Q$ (solid lines) showing that $LQ$ favors the clusterization into classes for higher $p$ while $Q$ almost always prefers the grouping into levels. Moreover, $p_c^Q$ is rather sensitive on the size of the network and converges to 0 as the network grows, while $p_c^{LQ}$ does not depend on the number of levels.}
\end{figure}

The crossing point $p_c$ depends on the number of levels and classes. Figure~\ref{pc} shows the change of $p_c$ upon variation of these two parameters with 2, 5 and 10 classes per level, respectively (from top to bottom). It can be seen that $p_c^{LQ}$ is higher than $p_c^{Q}$ for all values of levels and classes, and is by construction constant for a fixed number of classes per level. On the other hand, $p_c^Q$ strongly depends on network size which means that it favors different clusterizations as the number of levels increases, i.e., the lens of cluster detection becomes more coarse. Furthermore, it converges to 0 as $l$ grows, meaning that $Q$ favors the clusterization into levels for any $p \in [0,1]$, even though the classes on the same level are almost disconnected for small $p$.

\noindent These observations indicate that $LQ$ is more reliable than $Q$ to validate clusterizations in local cluster-connectivity networks. The discrepancies between the two measures originate from the fact that $Q$ compares the effective to the
expected fraction of links in the clusters, no matter if a link is \emph{possible} or not. The expected fraction of links is therefore underestimated in local cluster-connectivity networks, thus
the difference between expected and effective fraction of links (i.e., $Q$) is overestimated. On the other hand, $LQ$ only takes into account local link-expectations. 
Furthermore, note that modularity as high as 0.8 has been found in Erd\"os-R\'eni (ER) random graphs, scale-free networks and regular lattices \cite{Guimera:random, Massen:communities}. \\
%This arises at least partially from the fact that the summands of $Q$ depend on the rest of the network and therefore rely on \emph{global} information, like the number of nodes and links of the network. \\

In the last years, biological networks \cite{Grigorov} have attracted the attention of many scientists for their potential impact on the understanding of living systems. 
Metabolic and protein-protein interaction networks have been clustered 
by $Q$ optimization \cite{Amaral:cartography} and the MCL method \cite{Pereira}, respectively.
To investigate the behavior of $Q$ and $LQ$ on real-world networks we optimized the clusterizations of two recent realizations of the metabolic and protein-protein interaction networks of E. coli by simulated annealing (SA), using each of the two measures as cost function. For each temperature $T$, $c_1 n^2$ single-node and $c_2 n$ multi-node moves, like splitting and merging of (adjacent) communities, were performed, where $c_{1,2}$ are constants and $n$ is the number of nodes in the network. Furthermore, $T$ was iteratively reduced to $c_3 T$ with a constant $c_3<1$. This move set and cooling scheme is similar to the one used in \cite{Amaral:cartography}. The computational effort for the two measures scales as $O(K)$, even though the calculation of $LQ$ is slightly more expensive since it involves the determination of neighborhoods for each cluster.

\begin{figure}[ht!]
\begin{tabular}{c}
\includegraphics[angle=0,width=8.5cm]{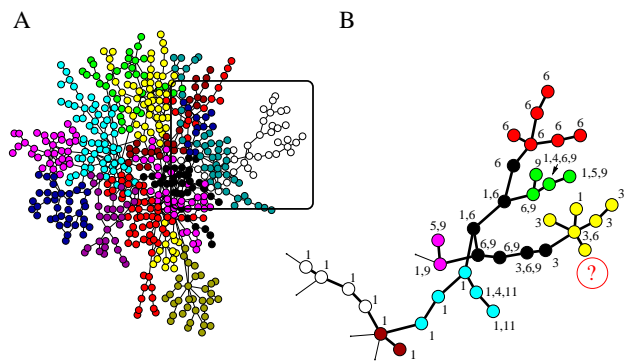}\\
\end{tabular}
\caption{\label{metnet} (Color online) {\bf (A)} Largest connected component of the metabolic network of \emph{E.coli}. The coloring scheme represents the clusterization found by optimizing modularity. Some colors are used twice. {\bf (B)} $LQ$ clusterization of the white $Q$ cluster with the annotation of different pathways. According to $LQ$ it is highly probable that the unassigned yellow node (N-Acetyl-alpha-D-glucosamine 1-phosphate, marked as ''?'') belongs to the carbohydrate metabolism (label 3).}
\end{figure}

(i) \emph{The metabolic network of E.coli.} We use the metabolic pathway database developed by Ma and Zeng \cite{MaZeng:met}, which has been derived from the Kyoto Encyclopedia of Genes and Genomes (KEGG) \cite{KEGG}. Figure~\ref{metnet} shows the largest connected component of the \emph{E.coli} metabolic network in this database. It contains 563 nodes and 708 links which have been treated undirected. Each node is assigned to between zero and nine out of 11 possible pathways. The optimization with fitness function $Q$ leads to a division into 16 clusters consisting of 35 metabolites on average (as colored in Figure~\ref{metnet}) and takes a value as high as $Q_{max}=0.82$. On the other hand, $LQ$ optimization leads to a maximum of $LQ_{max}=12.1$ with 132 clusters, containing each an average of 4.3 metabolites. The optimization of the two measures finds clusters at a different level, which yields complementary information. As expected, $Q$ is based on a global view and depends on the size of the network. As a consequence, optimizing a network with more metabolites would lead to larger $Q$ clusters. This problem is likely to arise because, as more data become available, the network and its largest connected component will grow. On the other hand, $LQ$ finds the lowest-level modules, independent on the rest of the network. Still, a mayor motivation to find clusters is to obtain information about presumed pathways of non-annotated metabolites. Figure~\ref{metnet}B zooms into one of the $Q$ clusters (white) and shows the splitting into smaller $LQ$ clusters. The numbers indicate the respective pathway(s) of the nodes. Note that an $LQ$ cluster is not necessarily fully contained in a $Q$ cluster, i.e., a smaller (local) cluster may be only \emph{partially} contained in a larger one. In the considered cluster of Figure~\ref{metnet}B, the further division is justified because it results in more homogeneous subclusters. The yellow community, for instance, contains mainly nodes belonging to the carbohydrate metabolism pathway (label 3). According to this, the unassigned node (N-Acetyl-alpha-D-glucosamine 1-phosphate, labeled as ''?'' in Fig.~\ref{metnet}B) can also be classified in pathway 3 with a high confidence. This would have been impossible when considering the white cluster obtained by $Q$ whose nodes are assigned mainly to pathway 6 (Glycan biosynthesis and metabolism) and 1 (Amino-acid metabolism).

To obtain a more quantitative analysis, we compute the conditioned probability
\begin{eqnarray*}
P[i,j]=P\left[\pi(i) \cap \pi(j) \neq \emptyset \, | \,c(i)=c(j)\right]
\end{eqnarray*} 
that two nodes $i$ and $j$, lying in the same cluster $c$, share at least one pathway ($\pi$). For the $Q$ clusterization, this probability is $P_Q[i,j]=0.57$, while $P_{LQ}[i,j]=0.73$, reflecting the higher homogeneity of the $LQ$ clusters. Comparison to the null-case, where nodes are picked at random from the network, yields $P_R[i,j]=0.26$ and the probability that any pair of linked nodes shares a pathway is 0.59, thus essentially the same as for the clustering with $Q$. 

\begin{figure}[ht!]
\begin{tabular}{c}
\includegraphics[angle=0,width=8.5cm]{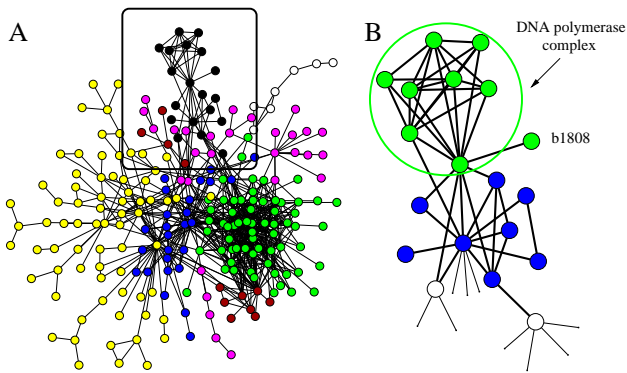}\\
\end{tabular}
\caption{\label{ppi} (Color online) {\bf (A)} Largest connected component of the PPI of \emph{E.coli}. The colors represent the clusterization found by optimizing modularity. {\bf (B)} $LQ$ clusterization of the black $Q$ cluster. The green circle contains proteins belonging to the DNA polymerase complex. The unknown protein b1808 is assigned to this complex according to $LQ$ while the complete $Q$ cluster is heterogeneous.}
\end{figure}

(ii) \emph{The protein-protein interaction (PPI) network of E.coli.} A set of 716 verified interactions involving 270 proteins of \emph{E.coli} has been reported \cite{Emili:ppi}. We again focused on the largest connected component consisting of 230 proteins and 695 undirected connections (Figure~\ref{ppi}). Identifying clusters can help to find indications about the function of unknown proteins. Again, modularity and localized modularity differ in the granularity of the clusters, similar to using two different lenses of a microscope. While the highest value for $Q$ has been found for a clusterization with 7 communities ($Q_{max}=0.49$), $LQ$ splits the network into 56 communities ($LQ_{max}=2.97$). An example where $LQ$ yields a more accurate ''guess'' is given in Figure~\ref{ppi}B, where the $LQ$ clusterization further subdivides the black cluster of Figure~\ref{ppi}A. The proteins in the green circle are part of the DNA polymerase complex (dnaE, dnaQ, dnaX, dnaQ, holA, holB, holC, holD and holE). According to $LQ$, the unknown protein b1808 appears to be a protein of this complex. On the other hand, the black cluster obtained by $Q$ is more heterogeneous which makes a functional assignment of b1808 difficult.

In conclusion, a new measure for the quality of network-clusterizations, called \emph{localized modularity}, has been introduced and compared to the widely used \emph{modularity}. Both measures can be used essentially in the same way.
The latter has been applied previously by others to assess the clusterization quality in many networks and has been used to find the best split of a dendogram and as fitness function in optimization algorithms. Finding clusters by optimizing a given fitness function has the advantage of not using any parameters (unlike many other clustering methods \cite{Reichardt:fuzzy,Enright,Palla:uncovering}). $Q$ depends on global properties like the network size and the cluster-connectivity. However, in many real-world networks, communities are merely connected locally, i.e., most pairs of clusters are not linked. We have called such organization \emph{local cluster-connectivity}.
By detailed investigation of model networks as well as the optimization of $Q$ and $LQ$ on two biological networks, we have provided evidence that the two measures give a view of different depth into the cluster structure. In contrast to $Q$, $LQ$ takes into account individual clusters and their nearest neighbors, generating high-confident clusters, irrespective of the rest of the network. Thus, the two measures provide complementary information. 
Furthermore, the $LQ$ approach can be generalized to 2$^{nd}$ or higher nearest neighbors which, albeit computationally more expensive, might yield additional insights, as if one were to use different lenses of a microscope.\\
This work was supported by a grant from the Swiss National Science Foundation.
%\newpage
{\footnotesize
\bibliography{fitness.bib}
}

\end{document}